\documentstyle[aps,prl]{revtex}
\def\la{{\langle}}
\def\ra{{\rangle}}

\newcommand{\beq}{\begin{equation}}
\newcommand{\eeq}{\end{equation}}
\newcommand{\beqa}{\begin{eqnarray}}
\newcommand{\eeqa}{\end{eqnarray}}

\newcommand{\wh}{\widehat}

\draft
\input{epsf}
\begin{document}
\epsfverbosetrue

\twocolumn[\hsize\textwidth\columnwidth\hsize
\csname @twocolumnfalse\endcsname

%      Title page and abstract...
\title{Time-of-arrival distribution for arbitrary potentials and 
Wigner's time-energy uncertainty relation.}
\author{A. D. Baute, R. Sala Mayato, J. P. Palao and J. G. Muga} 
\address{Departamento de F\'{\i}sica Fundamental II,
 Universidad de La Laguna,
La Laguna, Tenerife, Spain}
\author{I. L. Egusquiza}
\address{Fisika Teorikoaren Saila,
Euskal Herriko Unibertsitatea,
644 P.K., 48080 Bilbao, Spain} 
%\date{\today}

\maketitle

\begin{abstract}
A realization of the concept of ``crossing state'' invoked,
but not implemented, by Wigner, allows to advance in two 
important aspects of the formalization of the time of arrival 
in quantum mechanics: (i) For free motion, we find that the limitations
described by Aharonov et al. in
Phys. Rev. A 57, 4130 (1998)
for the time-of-arrival uncertainty at low energies
for certain measurement models are in fact already present in the
intrinsic time-of-arrival distribution of Kijowski;
(ii) We have also found a covariant generalization of this
distribution for arbitrary potentials and positions.
\end{abstract}   

\pacs{PACS: 03.65.-w}
\hfill 
]

In spite of the emphasis of quantum theory on the concept of
``observable'', the formalization of ``time observables'' is still a major
open and challenging question. The ``arrival time'' has in particular 
received much attention in the last few years 
(see \cite{MSP} for a recent review).     
Considering several candidates proposed for the time-of-arrival 
distribution in the simple free motion, one
dimensional case, 
some of us have recently argued \cite{MLP} in favor 
of a distribution originally proposed by Kijowski \cite{Ki}, $\Pi_K$,
because it satisfies a number of desirable conditions. 

This distribution can be associated with a POVM 
and obtained in terms of the eigenfunctions 
$|T,\alpha;X\ra$ ($\alpha=\pm$) of the 
time-of-arrival operator $\wh{T}$ introduced by Aharonov and Bohm
\cite{AB}
\beqa
\label{tiab}
&&\wh{T}=-\frac{m}{2}[(\wh{q}-X)\frac{1}{\wh{p}}
+\frac{1}{\wh{p}}(\wh{q}-X)]\,,
\\
\label{PiK}
&&\Pi_K(t+T;X;\psi(t))=\sum_\alpha |\la T,\alpha;X|\psi(t)\ra|^2\,,
\eeqa
(We consider here the general 1D case with both positive and 
negative momenta as in \cite{MLP,GRT,Gia}.)
where $m$ is the mass, $X$ is the arrival point, $\wh{q}$ and 
$\wh{p}$ are position and momentum operators, and  
\beq\label{tas0}
|T,\alpha;X\ra=e^{i\wh{H}_0 T/\hbar}(|\wh{p}|/m)^{1/2}
\Theta(\alpha \wh{p})|X\ra\,.
\eeq
(The operator $|\wh{p}|^{1/2}$ is defined by its action on momentum 
plane waves, $|\wh{p}|^{1/2}|p\ra=|p|^{1/2}|p\ra$.)
$\Pi_K(t+T;X;\psi(t))$
represents the probability density of arriving at $X$, at the 
instant $T+t$, for a given wavepacket $\psi(t)$, and  $t$
is the parametric time that characterizes the evolution of the
state $\psi(t)$. (Typically one sets $t=0$ so that $T$ is the
``nominal arrival time''.) This distribution satisfies in
particular the important 
{\it covariance condition} under time translations,
$\Pi_K(t+T-t';X;\psi(t+t'))=\Pi_K(t+T;X;\psi(t))$.
For other properties see \cite{MLP,Ki,Werner}.
       
In this letter we study two major aspects of this distribution
that had not been addressed: ({\it i})
For states with positive or negative momenta we 
shall obtain the states of minimum 
time uncertainty (for given energy width), and find the same type 
of limitation pointed out by Aharonov et al.\cite{Aetal};
({\it ii}) we shall also generalize (\ref{PiK}) for arbitrary
potentials.  

To handle conveniently this two issues 
let us first elaborate on the form of $\Pi_K$.
For $\alpha=+$ the contribution in (\ref{PiK}) can be interpreted as
a quantum version of the positive flux
at the time $T+t$ due to right moving particles. 
Similarly, for $\alpha=-$, one has a quantum version of minus the
negative flux due to left moving particles, again a positive quantity.
Explicitly,
\beqa
\Pi_{K,\alpha}(t+T)&=&\la \psi(t+T)|(|\wh{p}|/m)^{1/2}
\Theta(\alpha \wh{p})\delta(\wh{q}-X)
\nonumber\\
&\times&\Theta(\alpha \wh{p})(|\wh{p}|/m)^{1/2}|\psi(t+T)\ra\,,
\label{qop}
\eeqa
where $\delta(\wh{q}-X)=|X\ra\la X|$.
 The positive operator in (\ref{qop}) 
corresponds to the classical dynamical variable 
\beq\label{class}
\delta(q-X)\frac{\alpha p}{m}\Theta(\alpha p)\,,
\eeq
whose average represents the modulus of the flux of particles
of the classical ensemble that arrive at $X$ {\it from one side} at
a given time. This connexion was pointed out by Delgado \cite{D,Dcom}.  
There are of course many possible quantizations  of this quantity
but the symmetrical one in (\ref{qop}) turns out to be the only
one that satisfies the symmetry and minimum variance properties 
of $\Pi_K$.    

It is useful to write the positive operator in (\ref{qop})
in the form $|u_\alpha\ra\la u_\alpha|$,
where 
\beq\label{cs}
|u_\alpha\ra=|T=0,\alpha;X\ra
\eeq
is the ``crossing state''. As emphasized in \cite{MLP} it is
important to keep in mind that, being non-normalizable, its literal
interpretation is problematic. Only normalized wave packets peaked 
around these states have properties as close as desired to the sharp
crossing behaviour expected on intuitive grounds.      
 
Let us first discuss the point ({\it i}) related to 
the time-energy uncertainty principle. 
Since the Hamiltonian and the time operator (\ref{tiab})
are conjugate variables a
minimum uncertainty product can be established in the usual
fashion \cite{Gia}. 
However, Aharonov et al. have proposed, based on a series of 
models,  a second
 limitation on the possible values of the
time-of-arrival uncertainty:
$\delta t\, E_k > \hbar$,
where $\delta t$ is the width of the ``pointer variable'' used to
measure the arrival time, and $E_k$ ``the
typical initial kinetic energy of the particle'' \cite{Aetal}.
It is to be stressed that this relation is based on measurement models 
for the arrival time where some extra (clock) degree of freedom is  
coupled continuously with the particle.    
We shall see that in fact the ``intrinsic'' distribution
$\Pi_K$  (there is no explicit recourse to
additional pointer degrees of freedom to define $\Pi_K$) is
consistent with the behaviour that Aharonov et al. described for
their models \cite{Aetal}.

There are many other time-energy uncertainty relations
\cite{rev}, but  here we shall be mainly
interested in the version of  E. P. Wigner \cite{Wig}, because his
formalism is particularly suited for the
time of arrival. In the original paper Wigner did not consider
in detail any application and described a variational method to
find the states of minimum
uncertainty product, but did not actually obtain these states,
except in two analytically solvable cases. 
We shall extend Wigner's work in several 
directions by evaluating the states of minimum uncertainty 
product and applying the formalism to the arrival time. 
For completeness we shall next briefly summarize the main results
obtained by Wigner in \cite{Wig}, and add a number of comments and
observations relevant for our application. 
      
He defines the basic amplitude as
$\chi(t)=\la u|\psi(t)\ra$, where $|u\ra$ is in principle any state
vector. 
(Wigner's formalism encompasses many different time-energy
uncertainty products, depending on 
the $|u\ra$ chosen, each with its own physical interpretation.) 
Note that $t$
is considered the independent variable of $\chi$, and $|u\ra$ is fixed.
 $|u\ra$ is not necessarily 
a Hilbert space normalizable vector. It may also be, for example,  
a position, or a time-of-arrival eigenvector provided the minimum 
uncertainty product state
obtained is square integrable. 
$P(t)\equiv |\chi(t)|^2/\int_{-\infty}^{\infty}
|\chi(t)|^2 dt$ plays the role of a 
normalized distribution for being found at $|u\ra$ at time $t$.
This is not a standard 
quantity in the ordinary formulation of quantum mechanics (which assigns
probabilities only at fixed time $t$), but the
interpretation is consistent with the ordinary 
formulation in the following way:
Here ``being found'' implies operationally to perform measurements 
of $|u\ra\la u|$ at a given time $t$ for the members 
of an ensemble prepared at $t_0<t$ according to $\psi(t_0)$.
This is repeated at different times but the ensemble 
is always prepared anew at $t_0$ with the same specifications.
The distribution of positive counts as a function of 
time is proportional to $|\la u|\psi(t)\ra|^2$, and $P(t)$ is 
obtained when this distribution 
 can be normalized (which is not always the case).
($P(t)$ does not correspond to 
a continuous measurement that 
modifies $\psi(t)$.) The moments
of $P(t)$ are defined in the usual way, and in particular 
the {\it second moment with respect to } $t_0$ is defined as 
\beq\label{t2f}
\tau^2\equiv {\int_{-\infty}^{\infty} |\chi(t)|^2 (t-t_0)^2 dt}
/{\int_{-\infty}^{\infty} |\chi(t)|^2 dt}\,. 
\eeq
The information contained in $\chi(t)$ can also be encoded in 
its Fourier transform,   
\beq
\eta(E)\equiv h^{-1/2}\int \chi(t) e^{iEt/\hbar} dt.
\eeq
The conjugate variable $E$ has dimensions of energy, but 
$\eta(E)$ is not, in general, an energy amplitude in the conventional
sense. $\eta$ and the conventional energy amplitude of $\psi (t=0)$,
can be related by expanding $\psi(t)$ 
in a basis of energy eigenstates $|E,\alpha\ra$, 
\beq\label{etae}
\eta(E)=
h^{1/2}\sum_\alpha \la u|E,\alpha\ra\la E,\alpha|\psi(0)\ra
\Theta(E)\,.
\nonumber
\eeq
In a general case $\alpha$ is an index to account for the
possible degeneracy. 
In particular, for free motion, $\alpha=\pm$, and 
\beq
|E,\alpha\ra=\left(\frac{m}{2E}\right)^{1/4}|p=\alpha(2mE)^{1/2}\ra\,.
\eeq
Analogously to (\ref{t2f}) the 
second energy moment {\it with respect to $E_0$} is defined as 
\beq\label{subs}
\epsilon^2\equiv
{\int_0^\infty |\eta(E)|^2(E-E_0)^2 dE}/{\int_0^\infty
|\eta(E)|^2 dE}\,.
\eeq
Neither $t_0$ nor $E_0$ should in general be 
identified with the average values of time and
energy. They are instead reference parameters
fixed beforehand to evaluate the moments. As a consequence, $\epsilon^2$
and $\tau^2$ should not in general be identified with the
``variances'' $(\Delta E)^2$ and $(\Delta t)^2$,
which are the second moments with respect to
the average values.    
    
Since $\eta(E)$ and $\chi(t)$ are Fourier transforms of each other 
the uncertainty product $\epsilon\tau$ is greater than 
$\hbar/2$ (a peculiarity of 
time and energy with respect to position and momentum is that 
the equality is never satisfied \cite{Wig}).
In fact the bound increases substantially as $E_0$ decreases.     
Wigner sought for the function $\eta(E)$ that renders 
$\tau$ to a minimum for fixed $\epsilon$.    
In order to have a finite second moment $\tau^2$, $\eta(E)$ must 
vanish at the origin,  
$\eta(0)=0$, so $\eta$ must vanish at both ends of integration 
$E=0,\infty$. Using partial integration, and the notation 
$\eta_0=\eta e^{-iEt_0/\hbar}$
(Wigner showed that the minimum of $\tau\epsilon$ must correspond to a
real $\eta_0$.), one finds that  
\beq\label{t2}
\tau^2=\hbar^2 {\int_0^\infty |\partial \eta_0(E)/\partial E|^2 dE}/
{\int_0^\infty |\eta_0(E)|^2 dE}\,.
\eeq
The product $\tau^2\epsilon^2$ subject to the 
constraint of fixed $\epsilon^2$ is minimized by variational
calculus. This leads, using (\ref{subs}) and (\ref{t2}), to 
\beq\label{edeta}
-\hbar^2\frac{\partial^2\eta_0}{\partial E^2}+
\frac{\lambda'}{\epsilon^2}(E-E_0)^2\eta_0=(\tau^2+\lambda')\eta_0\,,
\eeq
where $\lambda'$ is a Lagrange multiplier.
This equation is formally similar to the Sch\"rodinger 
equation for the harmonic oscillator, except for the boundary condition 
at $E=0$, $\eta_0(0)=0$, and the subsidiary condition for $\epsilon$
(\ref{subs}). 
The minimum $\tau$ is obtained from the lowest eigenvalue
corresponding to the value of $\lambda'$ where the subsidiary condition 
is satisfied. Fortunately the solution depends only on the 
ratio $E_0/\epsilon$, namely 
 $\eta(E;E_0;\epsilon)=g(E/\epsilon;E_0/\epsilon)$, 
where $g$ is the solution of (\ref{edeta}) with $\epsilon=1$.  
Note also that, since $|\eta_0|^2=|\eta|^2$, the value of 
$t_0$ does not play any role in the minimization process. (Physically the 
state $\chi_{0}(t)$ corresponding to $t_0=0$ is 
valid for any other time $t_0$ by a shift of the argument.) 
The minimization of $\tau$ for fixed $E_0/\epsilon$ 
requires a method to solve the differential equation (\ref{edeta}) 
for many different values of $\lambda'$, until  
 $\epsilon^2=1$ is satisfied. 
In our calculation the successive values $\lambda'(n)$ have been
obtained with a Newton-Raphson algorithm,  
and the lowest eigenstate and eigenvalue of (\ref{edeta})
for each $\lambda'(n)$ with 
a very efficient ``relaxation method''
\cite{rel}.     

The only explicit case considered in the original paper by Wigner was
$|u\ra=|X\ra$. The corresponding $P(t)$
provides a time distribution 
for the {\it presence} of the particle at $X$ but not for its arrival.     
In an intriguing ``general observation'', Wigner stated that, instead     
of asking for the probability that the particle be at a definite landmark
in space, just at the time $t$, ``it would be more natural to ask ... for 
the probability that the particle crosses the aforementioned landmark
at the time $t$ from the left, and also that it crosses the landmark,
at a given time, from the right.'' But the paragraph ends 
 with  ``This point ..., interesting though it
may be, will not 
be elaborated further''. Precisely, it is our aim here to
elaborate further this question. Indeed the
probabilities mentioned by Wigner can be obtained by 
means of the crossing states $|u_+\ra$ and $|u_-\ra$
discussed before, see (\ref{cs}).
Specializing to states having only positive/negative
momentum $\Pi_K=\Pi_{K,\alpha}(t)=|\la u_\alpha|\psi(t)\ra|^2$
provides the arrival time distribution at $t$.
Considering $\chi(t)=\la u_\alpha|\psi(t)\ra$, 
we see that Wigner's probability density is nothing but 
Kijowski's distribution, $P(t)=\Pi_{K,\alpha}(t)=|\chi(t)|^2$.
Moreover, the Fourier transform
of $\chi(t)$ 
is in this case the standard energy amplitude,
$\eta(E)=\la E,\alpha|\psi(0)\ra$,   
so that  $\epsilon$ becomes the spread (around $E_0$) of
the ordinary energy distribution.  
\begin{figure}[h]
\centering
\epsfxsize=8cm
\epsfbox{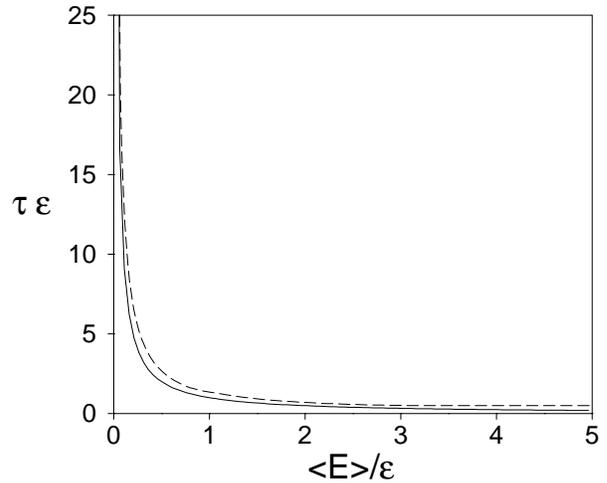}
\caption{\small $\epsilon\tau$ (in atomic units, $\hbar=1$)
versus $\la E\ra/\epsilon$ for the states 
of minimized uncertainty product (dashed line);  $\epsilon/\la E\ra$
(solid line).}  
\end{figure}
For a given ratio $E_0/\epsilon$ there is a minimum value of
$\tau\epsilon$. The family of states of minimal uncertainty 
$\eta(E;E_0;\epsilon)$ with the same 
ratio $E_0/\epsilon$ have in common the same value
of $\la E\ra/\epsilon$. Figure 1 shows $\tau\epsilon$
versus $\la E\ra/\epsilon$
for the states of minimized time--energy uncertainty product.
For comparison we also show the curve $\epsilon/\la E\ra$.
Clearly 
\beq\label{rela}
\tau > \hbar/\la E\ra\,,
\eeq 
which has the same {\it form} as the relation  proposed by
Aharonov et al. based on measurement models \cite{Aetal}.
However, $\tau$ is not due to the effect of any measuring apparatus,
it is an intrinsic uncertainty associated with an intrinsic 
time-of-arrival distribution. It is not the coupling introduced in 
a measurement between the particle and other degrees of freedom that
leads to this relation but the very quantum mechanical nature of the
particle alone and the lower bound of the energy.    
 
To elaborate the figure, $E_0/\epsilon$ has been increased regularly  
from the minimum possible value $-1$. (For smaller values it is
impossible to satisfy the subsidiary condition.)
For each value the minimization of $\tau\epsilon$ is 
performed and the corresponding  $\la E\ra/\epsilon$
is obtained.
As $\la E\ra/\epsilon\to\infty$ the minimum uncertainty
product tends to the (global) minimum $\hbar/2$, the same value found for
position and momentum, because the effect of the lower bound of the energy 
tends to disappear in that limit, and $\eta(E)$ becomes closer and
closer to a Gaussian centered at $E_0$ with variance
$\epsilon^2$ \cite{Wig}. However, in the opposite limit the lower bound
at $E=0$ plays an important 
role. Indeed, $\la E\ra/\epsilon\to 0$ corresponds to the limit 
$E_0/\epsilon\to -1$,
and the only way to satisfy the constraint is by  
strictly localizing $\eta(E)$ at $E=0$, ($\Delta E\to 0$),
but this completely delocalizes the conjugate time
variable, namely $\tau\to\infty$. Thus, (\ref{rela}) appears as a
consequence of the ordinary uncertainty principle, due to the tendency
of the minimum uncertainty product states to have smaller variances for
smaller energies.      
(The precise dependence for arbitrary values of $E_0/\epsilon$ has
to be obtained numerically.)  

The second question we shall address is the generalization of
the free motion distribution (\ref{PiK})
for arbitrary potentials and positions. 
 A generalization based on a quantization of classical 
expressions as in (\ref{tiab})
is problematic: the classical expressions for the time of arrival 
will rarely be analytical, not all phase space points lead to arrival, 
and the ordering problems may be formidable. The way out though,
will be surprisingly simple in terms of crossing states. 

There is in fact nothing that limits  (\ref{class}),
and the corresponding operator in (\ref{qop})
to free motion. In particular, the classical 
motivation for considering $|u_\alpha\ra$ a ``crossing state'' is
equally valid 
when an arbitrary  potential is present.      
The state (\ref{tas0}) may be regarded as one that has  evolved
``backwards'' with $H_0$ a time $T$ from the crossing state $|u_\alpha\ra$ 
 so that it becomes $|u_\alpha\ra$ precisely at
the nominal arrival time $T$.  
In the same vein we construct for an arbitrary Hamiltonian $H$  
\beq
|T,\alpha;X\ra=e^{i\wh{H}T/\hbar}|u_\alpha\ra\,,
\eeq
so that at the nominal arrival time $T$, $|u_\alpha\ra$
is recovered.
The generalization of the arrival time distribution for arbitrary
potentials is therefore 
\beq\label{Pi} 
\Pi(t+T;X;\psi(t))=\sum_\alpha
|\la u_\alpha|e^{-i\wh{H}T/\hbar}|\psi(t)\ra|^2\,.    
\eeq
\begin{figure}[h]
\centering
\epsfxsize=8cm
\epsfbox{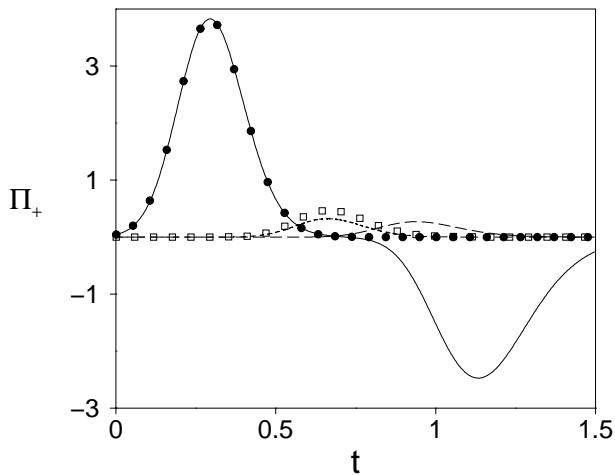}
\caption{\small Current density and $\Pi_+(t)$ 
before the barrier ($X=8$, solid line and 
circles), at the barrier ($X=12.25$, dashed line and squares),
and after the barrier ($X=15$, long dashed line --in
this case the two functions are
indistinguishable in the scale of the figure--),
for a collision of a Gaussian 
state (with minimum position-momentum uncertainty
product at $t=0$ and $\la \wh{q}\,\ra=\la \wh{p}\,\ra=5$,
 $\la (\wh{q}-5)^2\ra=1$) 
of a particle 
with mass  $m=1/2$ versus a square barrier of energy $40$
located between $x=12$ and $12.5$. All quantities in atomic units.} 
\end{figure}
It is evidently covariant under time translations as it should;  
in general it is not normalized, and it may be unnormalizable
(its classical counterpart shares these properties). 
For example, it may be constant for stationary states,
or periodic for oscillating coherent states in a harmonic potential,
but it provides in any case relative information by comparison of
two times. Consistent with this, the states $|T,\alpha;X\ra$ do not
form in general a complete basis.  
According to its classical analog 
it takes into account any crossings (not only first, or last).     
To illustrate the distribution   
we have evaluated $\Pi_+$ for a collision with tunnelling
at three different positions 
before, in, and after a square barrier, see  Figure 2. 
The only previous attempt to generalize $\Pi_K$ applied only
to asymptotic positions where the motion is essentially free
\cite{DM}.
Our generalization is instead valid for arbitrary positions.

One of us (J.G.M.)
acknowledges C. R. Leavens for helpful discussions.
This work was supported by Gobierno Aut\'onomo de Canarias 
(PB/95), MEC (PB97-1482), and CERION.   


\begin{references}
\bibitem{Dcom} The approach in 
\cite{D} is only physically applicable to states with positive or
negative momenta so that $\Pi_K$ could be interpreted as
a quantum version of the modulus of the flux. In the general case 
(\ref{PiK}) with contributions from the two subspaces this 
interpretation is only possible for each of the contributions 
separately but not for the sum.
\end{references}
\end{document}